-------------------------------------------------------------------------------

\documentstyle[11pt]{article}
\title{THE ELECTROWEAK CHIRAL LAGRANGIAN
                AND NEW PRECISION MEASUREMENTS}
\author{Thomas Appelquist and Guo-Hong Wu\\
Department of Physics, Yale University, New Haven, CT 06520}
\begin{document}
\setlength{\baselineskip}{24pt}
\maketitle

\begin{abstract}
A revised and complete list of the electroweak chiral lagrangian operators up
to dimension-four is provided. The connection of these operators to the $S$,
$T$ and $U$
parameters and the parameters describing  the triple gauge boson vertices
 $WW\gamma$ and $WWZ$  is made, and the size of these parameters
from new heavy physics is estimated using a one flavor-doublet model of
 heavy fermions. The coefficients of the chiral lagrangian operators
 are also computed in this model.
\end{abstract}

\section{Introduction}

  If electroweak symmetry breaking is driven by new strong interactions at TeV
energies, the electroweak chiral Lagrangian provides the most economical
description of electroweak physics below
this scale \cite{AB,Long,ASA}. Deviations from the standard (but Higgsless)
theory can be parameterized in terms of a low energy expansion, consisting of
operators of increasing dimension. All
deviations consistent with the $SU(2)_{L} \times U(1)$ symmetry are
describable in this way, with the level of accuracy depending on the the order
to which one goes in the low energy expansion.

   The rapidly improving precision of electroweak measurements
\cite{ewprecision} and the coming of LEP-II and possibly even higher energy
$e^{+}e^{-}$ colliders have focused new attention on the chiral lagrangian
approach [5-12].
 The purpose of this paper is to contribute to this
program in three ways. First, we provide a revised and complete list of all the
chiral lagrangian operators up to a certain order in the low energy expansion.
This list will include both CP-invariant and CP-violating operators. We then
provide a dictionary connecting these operators to the measured $S$, $T$ and
$U$
parameters and to the parameters describing the $W^{+} W^{-}\gamma$ and $
W^{+} W^{-}Z$ vertices. Finally, we report the results of a one-loop
estimate
of these operators arising from the presence of new fermions such as might be
present in technicolor type theories. The computation will be restricted to
 the CP conserving operators.

\section{The Chiral Lagrangian}

   The terms in the chiral lagrangian must respect the (spontaneously broken)
$SU(2)_{L} \times U(1)$ gauge symmetry. Experiment demands that the Higgs
sector also approximately respect a larger, $SU(2)_L \times SU(2)_C$ symmetry,
though the $SU(2)_C$ custodial symmetry is broken by the
Yukawa couplings and the $U(1)$ gauge couplings.
The possibility of an even larger global symmetry of the Higgs sector, leading
to the presence of pseudo-Goldstone bosons, will not be considered here.
The chiral lagrangian is thus constructed using the dimensionless unitary
unimodular matrix field $U(x)$, which transforms under $SU(2)_L \times
SU(2)_C$ as $(2,2)$. The covariant derivative of $U(x)$ is:
\begin{equation}
D_{\mu}U = \partial_{\mu} U + i g \frac{\vec{\tau}}{2} \cdot \vec{W_{\mu}}
U -i g' U \frac{\tau_3}{2} B_{\mu}.
\end{equation}

  In constructing the most general chiral $SU(2)_{L} \times U(1)_{Y}$
invariant effective lagrangian order by order in the energy expansion, it is
convenient to define the basic building blocks which are $SU(2)_{L}$ covariant
and $U(1)_{Y}$ invariant as follows:
\begin{eqnarray}
T & \equiv & U \tau_{3} U^{\dag}\ ,\; \; \; \; \;\;\;\;\;\;\;\;\;\;\;\;\;\;\;
V_{\mu} \;\, \equiv \;\, (D_{\mu} U ) U^{\dag} \\
W_{\mu\nu} & \equiv & \partial_{\mu} W_{\nu} - \partial_{\nu} W_{\mu} +
ig[W_{\mu}, W_{\nu}]
\end{eqnarray}
where $T$\,,\,$V_{\mu}$ and $W_{\mu\nu}$ have dimensions zero, one, and two
respectively.

   The familiar pieces of the chiral lagrangian, that
emerge for example from the $M_{H} \rightarrow \infty$ limit of the linear
theory at tree level, are:

\begin{equation}
{\cal L}_{0} \equiv \frac{1}{4} f^2 Tr[(D_{\mu}U)^{\dag}(D^{\mu}U)]
- \frac{1}{4} B_{\mu\nu}B^{\mu\nu} - \frac{1}{2} TrW_{\mu\nu}W^{\mu\nu},
\end{equation}
where $f \simeq 250$GeV is the symmetry breaking scale,
and $B_{\mu\nu} \equiv \partial_{\mu} B_{\nu} - \partial_{\nu} B_{\mu}$.
 The first term has dimension two, while
the second two (kinetic energy) terms have dimension four.
 The gauge  couplings to the quarks and leptons must also be added to Eq. 4.
 The Yukawa couplings of the quarks and leptons to the symmetry breaking sector
will be neglected here.

  There is one, additional dimension-two operator allowed by the
$SU(2)_{L} \times U(1)$ symmetry \cite{Long}:

\begin{equation}
{\cal L}_{1}^{\ '} \equiv  \frac{1}{4} {\beta}_1 g^{2} f^{2}
[Tr(TV_{\mu})]^{2}.
\end{equation}
This term, which does not emerge from the $M_{H} \rightarrow \infty$
limit of the renormalizable theory at tree level, violates the $SU(2)_{C}$
custodial symmetry even in the absence of the
gauge couplings. It is the low energy description of whatever
custodial-symmetry breaking physics exists, and has been integrated out, at
energies above roughly $\Lambda_{\chi} \equiv 4{\pi}f \simeq 3$TeV. In
technicolor theories, this breaking arises from the extended technicolor
interactions, typically at scales well above $\Lambda_{\chi}$.
 Used at tree level, ${\cal L}_{1}^{\ '}$
contributes directly to the deviation of the $\rho$ parameter from unity.

  At the dimension-four level, there are a variety of new operators that can be
written down. Making use of the equations of motion, and first
 restricting attention
to CP-invariant operators, the list can be reduced to eleven independent terms:

\begin{eqnarray}
{\cal L}_{1} & \equiv & \frac{1}{2}\alpha_1 gg' B_{\mu\nu} Tr(TW^{\mu\nu})\;\;
\;\;\;\;\;\;\;\;\;\;\;\;\;\;\;\;\;\;\;\;\;
{\cal L}_{2}  \equiv \frac{1}{2} i \alpha_2 g' B_{\mu\nu} Tr(T[V^{\mu},
 V^{\nu}]) \nonumber  \\
{\cal L}_{3} & \equiv & i \alpha_3 g  Tr(W_{\mu\nu}[V^{\mu}, V^{\nu}])  \;\;
\;\;\;\;\;\;\;\;\;\;\;\;\;\;\;\;\;\;\;\;\;
{\cal L}_{4}  \equiv  \alpha_4 [Tr(V_{\mu}V_{\nu})]^2  \nonumber \\
{\cal L}_{5} & \equiv & \alpha_5 [Tr(V_{\mu} V^{\mu})]^2  \;\;
\;\;\;\;\;\;\;\;\;\;\;\;\;\;\;\;\;\;\;\;\;\;\;\;
{\cal L}_{6}  \equiv  \alpha_6 Tr(V_{\mu} V_{\nu})Tr(TV^{\mu})
Tr(TV^{\nu}) \nonumber \\
{\cal L}_{7} & \equiv & \alpha_7 \  Tr(V_{\mu} V^{\mu})Tr(TV_{\nu})
Tr(TV^{\nu})\;\;\;\;\;\;\;
{\cal L}_{8}  \equiv  \frac{1}{4} \alpha_8 \  g^2 \ [Tr(TW_{\mu\nu})]^2
\nonumber \\
{\cal L}_{9} & \equiv & \frac{1}{2} i \alpha_9 g Tr(TW_{\mu\nu})Tr(T[V^{\mu},
V^{\nu}])\;\;\;\;\;\;\;\;\;\;
{\cal L}_{10}  \equiv \frac{1}{2} \alpha_{10} [Tr(TV_{\mu})Tr(TV_{\nu})]^2
 \nonumber \\
{\cal L}_{11} & \equiv & \alpha_{11} g {\epsilon}^{\ \mu\nu\rho\lambda}\
Tr(TV_{\mu})Tr(V_{\nu}W_{\rho\lambda})
\end{eqnarray}
The first ten terms were written down by Longhitano \cite{Long}. They have been
reconsidered recently by several authors \cite{Ruj,HeV}. The operator
${\cal L}_{11}$ is new \cite{Fer} and it completes the list of all $CP$
invariant operators up to dimension four (see Appendix~A for more
details). ${\cal L}_{11}$ corresponds to a CP-conserving, but $C$ and $P$
violating, term in the general parameterization of the triple gauge
boson vertex. It will be considered further in Section 4. We use the convention
${\epsilon}_{0123}=-{\epsilon}^{0123}=1$.

Longhitano's list \cite{Long} also contains CP-violating, dimension-four
operators. The full list of such operators, after making use of the equations
of motion, contains five terms in addition to
those written down by Longhitano. They are all listed in Appendix A. In this
paper, detailed considerations will be restricted to the $CP$ invariant
operators.

\section{Oblique Corrections}

  Since experimental work is so far restricted to energies below the W-pair
threshold, the only operators in the above list that have been
directly constrained experimentally are those that contribute to the gauge
boson two-point functions. In addition to ${\cal L}_{0}$, they
 are ${\cal L}_{1}^{'}$, ${\cal L}_{1}$ and ${\cal L}_{8}$, and they
 can be directly related
to the $S$, $T$ and $U$ parameters introduced by Peskin and Takeuchi
 \cite{PeT}. By setting the Goldstone boson fields to zero in these
operators (``going to unitary gauge"), one finds
\begin{eqnarray}
S & \equiv & - 16 \pi \frac{d}{dq^2}{\Pi}_{3B}(q^2){|}_{q^2=0}  =  -16 \pi
{\alpha}_1 ,\\
\alpha T & \equiv & \frac{e^2}{c^2s^2m_{Z}^2} ({\Pi}_{11}(0) - {\Pi}_{33}(0))
 =  2 g^2 \beta_1,\\
U & \equiv & 16 \pi \frac{d}{dq^2}[{\Pi}_{11}(q^2) - {\Pi}_{33}(q^2)]
{|}_{q^2=0}  =  -16 \pi {\alpha}_8.
\end{eqnarray}
The $\Delta\rho( \equiv \rho -1)$ parameter is related to $T$ by
 $\Delta\rho_{new} = \Delta\rho
- \Delta\rho_{SM} = \alpha T$, where $\Delta\rho_{SM}$ is the contribution
arising from standard model corrections.

\section{The Triple Gauge Vertex}

  The next generation of $e^{+}e^{-}$ colliders will operate above the W pair
production threshold, and will therefore be able to directly measure the triple
gauge vertices (TGV's).
The most general polynomial structure of the TGV has been derived \cite{Hag2}
by
imposing Lorentz invariance and on-shell conditions for the $W^{+}$ and
$W^{-}$. The corresponding effective lagrangian for this vertex is:
\begin{eqnarray}
 {{\cal L}_{WWV} \over g_{WWV}} & = &  i g_1^{V}(W^{+}_{\mu\nu}W^{-\mu}V^{\nu}
 -  W^{-}_{\mu\nu}W^{+\mu}V^{\nu})
 + i {\kappa}_{V}W^{+}_{\mu}W^{-}_{\nu} V^{\mu\nu}  \nonumber \\
 & & +\, \frac{i{\lambda}_{V}}{\Lambda_{\chi}^2}
 W^{+}_{\mu\nu} W^{-\nu}_{\ \ \ \rho} V^{\rho\mu}
 - g_4^{V}W^{+}_{\mu}W^{-}_{\nu}
 ({\partial}^{\mu} V^{\nu} + {\partial}^{\nu} V^{\mu} )  \nonumber \\
 & & +\, g_{5}^{V} {\epsilon}^{\mu\nu\rho\lambda}
 [\  W^{+}_{\mu} ({\partial}_{\rho}W^{-}_{\nu}) - ({\partial}_{\rho}
 W^{+}_{\mu}) W^{-}_{\nu}\  ]V_{\lambda}
 + i {\tilde{\kappa}}_{V} W^{+}_{\mu}
 W^{-}_{\nu} {\tilde{V}}^{\mu\nu} \nonumber \\
 & & +\, \frac{i {\tilde{\lambda}}_{V} }{\Lambda_{\chi}^2}
 W^{+}_{\mu\nu} W^{-\nu}_{\ \ \ \rho} {\tilde{V}}^{\rho\mu},
\end{eqnarray}
where $V=\gamma$ or $Z$, $\ W^{\pm}_{\mu\nu}={\partial}_{\mu}W^{\pm}_{\nu} -
{\partial}_{\nu}W^{\pm}_{\mu}$, $\ V_{\mu\nu}={\partial}_{\mu}V_{\nu} -
{\partial}_{\nu}V_{\mu}$, and ${\tilde{V}}_{\mu\nu}=\frac{1}{2}
{\epsilon}_{\mu\nu\rho\lambda} V^{\rho\lambda}$. The coupling constants
$g_{WW\gamma}$ and $g_{WWZ}$  are
given by $g_{WW\gamma}= -e$ and $g_{WWZ}= -e \frac{c}{s}$, where $e$ is
the renormalized electric charge, and $c \equiv \cos \theta_{w}{|}_{Z}$ and
 $s \equiv \sin \theta_{w}{|}_{Z}$ are the ``Z-standard"
definition of the weak mixing angle \cite{Pes}:
\begin{eqnarray}
s^2 c^2 \equiv \frac{\pi \alpha}{\sqrt{2} G_F m_Z^2},
\end{eqnarray}
 the value of $\alpha$ in this definition is  $\alpha^{-1} (m_Z^2)
= 128.80 \pm 0.12$ \cite{PeT}.

This effective
lagrangian contains two terms of dimension six. Unlike
Ref.\cite{Hag2}, we have used
the scale $\Lambda_{\chi} \equiv 4\pi f$ to define the corresponding
dimensionless parameters  ${\lambda_V}$ and ${{\tilde{\lambda}}_V}$
in these terms. This is the natural thing to do if
this effective lagrangian is assumed to arise from integrating out only
 high energy
physics, above the scale $\Lambda_{\chi}$. That is what is
being done here, since we want to make direct contact between this effective
lagrangian and the terms in the chiral lagrangian in Section 2.
The above dimension-six terms then correspond to dimension-six operators
in the chiral lagrangian, which are higher order in the low energy expansion,
(suppressed by a factor of ${\cal O} ( \frac{p^2}{\Lambda_{\chi}^2} ) $
relative to the dimension-four operators).  These dimension-six terms will
be neglected from here~on.

\newpage

{}~

{}~

{}~

{}~

   Fig.~1   Convention for the triple gauge boson vertex, with $V= \gamma$
 or $Z$.

\vskip 0.2in

   Using the convention displayed in the diagram above, the
Feynman rules for the triple gauge vertices, described by the above
effective lagrangian, can be written down:
\begin{eqnarray}
 {\Gamma}_{V}^{\mu\nu\rho}(p,q,k)& =& g_1^{V}(p-q)^{\rho} g^{\mu\nu}
 + (g_1^V + \kappa_V ) ( k^{\mu} g^{\rho\nu} - k^{\nu} g^{\rho\mu} )\nonumber
\\
 & &  +\, i g_4^{V} ( k^{\mu}g^{\rho\nu}+ k^{\nu} g^{\rho\mu} )
 + i g_5^{V} {\epsilon}^{\rho\mu\nu\lambda} ( p- q)_{\lambda}
 \nonumber \\
 & &\,  - {\tilde{\kappa}}_V {\epsilon}^{\rho\mu\nu\lambda} k_{\lambda}
\end{eqnarray}
where $k = p+q$  by conservation of momenta, and terms corresponding to
 dimension-six operators have been ignored.
 Note that the first two  terms in the above expression
 are $C$ and $P$ invariant (thus $CP$ invariant), the $g_5^V$ term is $CP$
invariant but $C$ and $P$ violating, and that the other two terms
are $CP$ violating.

 To make the connection between the TGV parameters and the general chiral
lagrangian of  Section 2, we restrict attention to the $CP$
conserving sector. We also use the convention of defining the parameters of
the TGV's to include the effects of corrections to the $W$, $Z$ and $\gamma$
propagators and effects of the $\gamma$ $Z$ mixing, coming  from physics
 above $\Lambda_{\chi}$. Thus
these parameters are related to those in the chiral lagrangian through the
latter's contribution to both gauge boson three-point and two-point functions.
Using the ``Z-standard" definition of the renormalized weak mixing angle
$\theta_{w}{|}_{Z}$  and
by going to the unitary gauge, one finds \cite{Hol2}:
\begin{eqnarray}
 g_1^Z - 1 & = & \frac{1}{s^2(c^2-s^2)}e^2 {\beta}_1
 + \frac{1}{c^2(c^2-s^2)}e^2 {\alpha}_1
 + \frac{1}{s^2c^2}e^2 {\alpha}_3  \nonumber \\
 g_1^{\gamma} - 1 & = & 0 \nonumber \\
{\kappa}_Z -1 & = & \frac{1}{s^2(c^2-s^2)}e^2 {\beta}_1
 + \frac{1}{c^2(c^2-s^2)} e^2 {\alpha}_1
 + \frac{1}{c^2} e^2 ( \alpha_1 - {\alpha}_2 )
 + \frac{1}{s^2} e^2 ({\alpha}_3 - {\alpha}_8 +{\alpha}_9) \nonumber  \\
 {\kappa}_{\gamma} - 1 & = &  \frac{1}{s^2} e^2 (- {\alpha}_1 + {\alpha}_2 +
 {\alpha}_3 - {\alpha}_8 + {\alpha}_9 ) \nonumber \\
g_5^Z & = & \frac{1}{s^2c^2} e^2 {\alpha}_{11} \nonumber \\
g_5^{\gamma} & = & 0,
\end{eqnarray}
 where $s \equiv \sin \theta_{w}{|}_{Z}$, $c \equiv \cos \theta_{w}{|}_{Z}$.

 The first two terms in the expressions for $g_1^Z-1$ and $\kappa_Z -1$ are the
contributions from the gauge boson two-point functions of the chiral
 lagrangian. The
$\beta_1$ dependence of the first term is a consequence of using the
``Z-standard"  definition $\theta_{w}{|}_{Z}$, whose renormalization
 depends on the
$Z$ mass renormalization induced by the $\beta_1$ term in the chiral
lagrangian. The other terms  in $g_1^Z-1$ and $\kappa_Z-1$ come from
the gauge boson three-point
 function contribution of the chiral lagrangian.  The parameters
$\kappa_{\gamma}-1$ and $g_5^Z$
 have contributions only from the three-point functions of the chiral
lagrangian.
Note that  $g_1^{\gamma}$ measures the electric charge
 of the $W$ in unit of $e$
and that the vanishing of $g_5^{\gamma}$ is a consequence
of $U(1)_{em}$ gauge invariance.

  The right hand side of Eq.(13) measures deviations  from the
 standard-model tree-level predictions, coming from new high energy physics.
 The $\alpha$ parameters are expected to be of order $\frac{1}{16{\pi}^2}$,
or smaller if they arise only from weak-isospin breaking effects.
 This expectation will be born out in the
 model computations in the next section. In order to isolate these effects
experimentally, one-loop radiative corrections within the standard model
(arising from momentum scales less than ${\Lambda}_{\chi}$~) must also be
 included. These corrections will not be considered in this paper.

  Notice  that the $g_5^Z$ term is a weak-isospin breaking operator and
vanishes in any  weak-isospin symmetric theory. Thus it
is not related  to the  Wess-Zumino-Witten anomaly \cite{WZW},
 which is not dependent on
weak-isospin symmetry breaking in the new fermionic sector.
 Since we will restrict
attention to the low energy effects of an anomaly free heavy fermion sector, we
will not need to worry about the WZW term.

\section{One-Loop Estimates}

   The electroweak chiral lagrangian operators have been connected to
the parameters ($S$, $T$ and $U$) that describe the ``oblique'' corrections
coming from new physics above $\Lambda_{\chi}$ (Section 3), and to
 terms in the TGV that will be directly probed at LEP-II (Section 4). We next
  estimate the size of these parameters  arising from physics above
 $\Lambda_{\chi}$ in a simple model consisting of one flavor-doublet of heavy
fermions $U$ and $D$.  By assigning both $U$ and $D$ to the fundamental
representation of an $SU(N)$ group, this model can be viewed as a simplified
version of a technicolor theory with the  technicolor interactions
neglected.

 A small weak-isospin asymmetry arising from the mass splitting in the fermion
doublet is also included.  The masses of U- and D-type fermions
 are denoted by $m_U$ and $m_D$ respectively, and
the weak-isospin asymmetry parameter is
defined by $\delta \equiv \frac{m_U - m_D}{m_U + m_D}$. The electric charges of
the U- and D-type fermions  are set to be $+\frac{1}{2}$ and $-\frac{1}{2}$
respectively by anomaly cancellation  condition for the new fermionic sector.

   The computation of the $S$, $T$ and $U$ parameters has been done before
 (see for example \cite{PeT}). The results are:
\begin{eqnarray}
S &=& \frac{N}{6 \pi}(1-Y\ln \frac{m_U^2}{m_D^2}) = \frac{N}{6 \pi}  \\
\alpha T & \simeq & \frac{Ne^2}{48 {\pi}^2 s^2 c^2} \frac{(\Delta m)^2}{m_Z^2}
 \\
U & \simeq & \frac{8N}{15 \pi} {\delta}^2
\end{eqnarray}
where $\Delta m \equiv m_U - m_D$, and the hypercharge $Y$ for each
left-handed U and D doublet is zero for anomaly cancellation.
Note that  since the $U$ and $D$ are the source of electroweak
symmetry breaking, $f \simeq 250$GeV can be expressed in terms of $m_U$ and
$m_D$. In our simple model, the one-technifermion-loop
expression for $f$ is $f^2 = \frac{N}{4 {\pi}^2} m^2 \ln \frac{\Lambda^2}
{m^2}$, where $m \simeq m_U \simeq m_D$ and where $\Lambda$ is an ultraviolet
 cutoff. Of course, in a real technicolor theory, the technifermion masses
will be soft, having values of order $m_U$ and $m_D$ at scales of order
$\Lambda_{\chi}$, and falling with increasing momentum. The integrals will
then be
cut off in the ultraviolet at momenta of order $\Lambda_{\chi}$, and the $\ln
\frac{\Lambda^2}{m^2}$ will be replaced by a factor of order unity.

  We turn next to the triple gauge boson vertices.
To simplify the computations, we consider the process
$e^+e^- \rightarrow W^+W^-$, with the $W$'s on mass-shell,
 that will be studied at LEP-II. Recall
 that the new fermion masses (of order ${\Lambda}_{\chi}$ ) are much larger
than the center of mass energy of the reaction and the $W$ mass.

   Dimensional regularization and on-shell renormalization are used
throughout the computation. As for ${\gamma}_5$, we employ in $n$ dimensions
the definition \cite{Cha}:
\begin{eqnarray}
  {\gamma}_{5}{\gamma}_{\sigma}+ {\gamma}_{\sigma}{\gamma}_{5}= 0
 \;\;\;\;\;\;\;\;\;\;\;\;
 {{\gamma}_{5}}^2 =1
\end{eqnarray}
where $\sigma = 0,1,\cdots, n-1$, and the prescription:
\begin{eqnarray}
tr{\gamma}_5 {\gamma}_{\mu} {\gamma}_{\nu} {\gamma}_{\rho} {\gamma}_{\lambda}
 = 4i {\epsilon}_{\mu\nu\rho\lambda}
\end{eqnarray}
for ${\mu},{\nu},{\rho},{\lambda}=0,1,2,3$. The resulting ambiguity associated
 with the trace of one ${\gamma}_{5}$ and six or more ${\gamma}_{\sigma}$
matrices is resolved by demanding that the
${\gamma}_{\sigma}$'s are not to be anticommutted with ${\gamma}_5$ inside the
trace.

{}~

{}~

{}~

{}~

            Fig. 2.   The (direct) one-loop  contributions to the $WWV$
($V= \gamma$ or $Z$)  vertex function coming from the heavy fermion doublet
 $U$ and $D$. The electric charge assignments are
 $Q_U=+\frac{1}{2}$ and $Q_D=-\frac{1}{2}$, for anomaly cancellation.

\vskip 0.2in

    The $WW{\gamma}$ vertex will be considered first.  The two
 Feynman diagrams  contributing directly to the three-point
 function are shown in Fig. 2. Indirect
 contributions coming from the gauge boson two-point functions must also be
 included. Keeping only terms linear in the external momenta (the higher order
 terms are suppressed at least by inverse square powers of
 ${\Lambda}_{\chi}$ ) and performing the electric charge and mass
 renormalization at the one-loop level, we find, to order $\delta^2$,
\begin{eqnarray}
\lefteqn{{\Gamma}_{\gamma}^{\mu\nu\rho} (p,q,p+q) = } \nonumber \\
 & & [- g_{\rho\mu}(2p+q)_{\nu} + g_{\rho\nu} (p+2q)_{\mu}
 + g_{\mu\nu} (p-q)_{\rho}] \cdot
  (1 - \frac{Ne^2}{96 {\pi}^2 s^2} ) \nonumber  \\
 & & - \, \frac{Ne^2}{96{\pi}^2s^2} [g_{\rho\mu} (p-q)_{\nu} + g_{\rho\nu}
(p-q)_{\mu} -g_{\mu\nu} (p-q)_{\rho} ] \nonumber \\
 & & + \, \frac{Ne^2}{96{\pi}^2s^2} \frac{4}{5} {\delta}^2
[g_{\rho\nu} (p+q)_{\mu} - g_{\rho\mu} (p+q)_{\nu} ]
\end{eqnarray}
where  $\delta =\frac{m_U - m_D}{m_U + m_D}$.

   By using the on-shell condition for the $W$'s, the above expression
 can be rewritten in the form of Eq. (12), and the
 values for the $WW\gamma$ vertex parameters can  be extracted:
\begin{eqnarray}
g_1^{\gamma} - 1 = 0 \;\;\;\;\;\;\;\;\;\;
{\kappa}_{\gamma} - 1 =  - \frac{Ne^2}{96{\pi}^2s^2} ( 1 - \frac{4}{5}
{\delta}^2 )  \;\;\;\;\;\;\; g_5^{\gamma} = 0.
\end{eqnarray}
The vanishing of $g_1^{\gamma} -1$ and $g_5^{\gamma}$ are consistent with
the chiral lagrangian analysis of Section 4.

  The one-loop computation for the $WWZ$ vertex can be  similarly done.
 Keeping only terms linear in the external momenta, performing
  the ``$Z$ charge", mass and weak mixing angle renormalization
 and using the ``Z-standard"
definition of the weak mixing angle $\theta_{w}{|}_{Z}$, it is found that,
\begin{eqnarray}
\lefteqn{{\Gamma}_{Z}^{\mu\nu\rho} (p,q,p+q) = } \nonumber \\
 & &  [- g_{\rho\mu}(2p+q)_{\nu} +
 g_{\rho\nu} (p+2q)_{\mu} + g_{\mu\nu} (p-q)_{\rho}] \cdot \nonumber \\
  & & \cdot  (1 - \frac{Ne^2}{96{\pi}^2s^2}
   - \frac{Ne^2}{96 {\pi}^2} \frac{1}{c^2 (c^2-s^2)}
   + \frac{Ne^2}{96 {\pi}^2} \frac{1}{s^2c^2(c^2-s^2)}
 \frac{(\Delta m)^2}{m_Z^2} ) \nonumber \\
 & & + \, \frac{Ne^2}{96{\pi}^2c^2}
 [g_{\rho\mu} (p-q)_{\nu} + g_{\rho\nu}
(p-q)_{\mu} -g_{\mu\nu} (p-q)_{\rho} ]
\cdot ( 1- \frac{1}{s^2} \frac{2}{5} \delta^2)       \nonumber \\
 & & +\, \frac{Ne^2}{96{\pi}^2s^2} \frac{4}{5} {\delta}^2
  [g_{\rho\nu} (p+q)_{\mu} - g_{\rho\mu}
 (p+q)_{\nu} ]     \nonumber \\
 & & +\, \frac{-iNe^2}{96{\pi}^2 c^2s^2}
 {\epsilon}_{\alpha\nu\rho\mu} (p-q)^{\alpha}(\delta + {\cal O} ({\delta}^3) ),
\end{eqnarray}
 where terms of higher order than $\delta^2$ have been dropped.

  By using the on-shell condition for the $W$'s,  the above result  can be put
into the form of Eq. (12). The one-loop values for the $WWZ$ vertex
 parameters  are:
\begin{eqnarray}
{g_1^Z}-1 & = &  - \frac{Ne^2}{96 {\pi}^2}\cdot ( \frac{1}{s^2 (c^2-s^2)}
- \frac{1}{s^2c^2(c^2-s^2)} \frac{(\Delta m)^2}{m_Z^2})
+ \frac{Ne^2}{96{\pi}^2c^2s^2} \frac{2}{5} \delta^2       \nonumber \\
{\kappa}_Z -1 & = &  - \frac{Ne^2}{96 {\pi}^2}
\frac{1-2s^2c^2}{s^2c^2(c^2-s^2)}
+ \frac{Ne^2}{96{\pi}^2} \frac{1}{s^2c^2(c^2-s^2)} \frac{(\Delta m)^2}{m_Z^2}
+ \frac{Ne^2}{96{\pi}^2s^2} \frac{4}{5} \delta^2        \nonumber  \\
{g_5^Z} & = &  \frac{Ne^2}{96 \pi^2} \frac{1}{s^2c^2}
(\delta + {\cal O} ({\delta}^{3}) )
\end{eqnarray}

  Recall that  the  $g_5^Z$ term is $CP$ conserving but
 parity and charge conjugation violating. This term is proportional to
the axial coupling of the neutral gauge boson to the fermions,
 and  is odd under the interchange
 of the masses of the $U$ and $D$ fermions.  This results in a linear
 dependence on $\Delta m \equiv m_U - m_D$ for $\Delta m \ll m_U \simeq m_D$.

   All the CP-conserving chiral lagrangian coefficients that enter the
gauge boson two- and three-point functions can now be determined from
 Eqs. (7), (8), (9) and (13). To order $\delta^2$, they are given by:
\begin{eqnarray}
\beta_1 & \simeq & \frac{N}{96 {\pi}^2 c^2} \frac{(\Delta m)^2}{m_Z^2}  \\
\alpha_1 & = & - \frac{N}{96 {\pi}^2 } \\
\alpha_2 & = & - \frac{N}{96 {\pi}^2} \\
\alpha_3 & = & - \frac{N}{96 {\pi}^2 } (1- \frac{2}{5} \delta^2 ) \\
\alpha_8 & = & - \frac{N}{96{\pi}^2} \frac{16}{5} {\delta}^2    \\
\alpha_9 & = & - \frac{N}{96{\pi}^2} \frac{14}{5} {\delta}^2    \\
\alpha_{11}& = &  \frac{N}{96 {\pi}^2} (\delta + {\cal O} ({\delta}^{3}) )
\end{eqnarray}
 Note that ${\alpha}_{11}$ is linear in the mass difference of the U and D
 fermions for a small weak-isospin asymmetry. We stress that these results
 apply only to our simple model in which technicolor interactions are
neglected. They will be modified by the strong interaction of a realistic
technicolor theory.

    All the $\alpha$ parameters are finite dimensionless constants,
not suppressed by powers of $\frac{1}{m^2}$, with $m \simeq m_U \simeq m_D$.
  This is consistent with the fact that
they are the coefficients of dimension-four  operators, in which the
heavy physics does not decouple.

  The size of the TGV parameters coming from physics above $\Lambda_{\chi}$
can be estimated in this simple model by assigning values for N and the
masses. We choose:
\begin{eqnarray}
N = 4, \;\;\;\;\;\;\;
m_U \simeq m_D  \simeq 1.5 TeV,
\end{eqnarray}
together with $\Delta m \equiv m_U - m_D  \ll \frac{m_U + m_D}{2}$.
It is then found that
\begin{eqnarray*}
 \kappa_{\gamma} -1 \simeq - 1.8 \times 10^{-3}.
\end{eqnarray*}

The values of  $g_1^Z-1$ and $\kappa_Z-1$  depend on the size of
$\beta_1$, and therefore on the mass splitting $\Delta m$ between the U and D
fermions.  The $g_5^Z$ parameter is proportional to $\Delta m$.
  Their values are tabulated below for various values of $\Delta m$.

\vskip 0.2in

Table 1. The size of the TGV parameters ($g_1^Z-1$, $\kappa_Z-1$ and $g_5^Z$)
 in the simple model being considered here. The results are for
 $N=4$ and $m_U\simeq m_D \simeq 1.5TeV$, with various values of $\Delta m$.
 These results will be modified by the strong interactions of a realistic
technicolor theory.

\begin{tabular}{|c|c|c|c|c|c|c|c|c|} \hline\hline
$\Delta m$ &
$\Delta\rho_{new}$ &
\multicolumn{3}{c|}{$g_1^Z-1$ $(\times 10^{-3})$} &
\multicolumn{3}{c|}{$\kappa_Z-1$ $(\times 10^{-3})$} & $g_5^Z$ \\ \cline{3-8}
 (GeV) & $(\times 10^{-2})$ &
 2-pt & 3-pt & total &
 2-pt & 3-pt & total & $(\times 10^{-4})$ \\ \hline
150 & 1.25  & 10.6 & -2.3 & 8.3  & 10.6 & -1.8  & 8.8  & 1.0  \\ \hline
100 & 0.56  & 4.2  & -2.3 & 1.9  & 4.2  & -1.8  & 2.4  & 0.7  \\ \hline
44  & 0.11  & 0.0  & -2.3 & -2.3 & 0.0  & -1.8  & -1.8 & 0.3  \\ \hline
0.0 & 0.0   & -1.0 & -2.3 & -3.3 & -1.0 & -1.8  & -2.8 & 0.0  \\ \hline \hline
\end{tabular}

\vskip 0.3in
We have separated the gauge boson two-point function contribution from the
 gauge boson three-point function contribution in $g_1^Z-1$ and $\kappa_Z-1$
 in order to see explicitly the dependence on $\beta_1$
 (or $\Delta \rho_{new}$). Recall that
 the gauge boson two-point function contributions are the first  two terms
in the expressions for $g_1^Z-1$ and $\kappa_Z-1$ in Eq. (13).

 It can be seen from the above table that for a large mass splitting
between  the U and D fermions, the gauge boson two-point
function contribution dominates and is opposite in sign to the
gauge boson three-point
function contribution. The overall signs for $g_1^Z~-~1$ and $\kappa_Z-1$ are
both positive.
 With the mass difference  about 45GeV, the gauge boson
two-point function contribution vanishes and only the gauge boson
three-point function contribution survives.  For an even smaller mass
splitting, the gauge boson two-point function
contributes with the same sign  as the gauge boson three-point function.

 Notice that the typical size of both $g_1^Z-1$ and $\kappa_Z-1$
in this simple model, coming from physics above $\Lambda_{\chi}$, is
   ${\cal O}(\pm 10^{-3})$ for a small to moderate mass splitting,
and that the size of $\kappa_{\gamma}-1$ is ${\cal O}(- 10^{-3})$.  In
models with more than one flavor-doublet (such as the one family technicolor
model), the size of the TGV parameters could
be as large as ${\cal O} (10^{-2})$. Assuming the strong technicolor
interactions do not substantially change these results, the
 estimates indicate that deviations of the size of the TGV parameters
 from standard model predictions could possibly be seen in
a 500 GeV $e^{+}e^{-}$ linear collider,
but are too small to be seen in LEP-II experiments~\cite{Ein}.

\section{Conclusions}

 The $S$, $T$ and $U$ parameters introduced by Peskin and Takeuchi give a
complete description of the oblique radiative corrections from physics
above ${\Lambda}_{\chi}$. As the energies of forthcoming  experiments
move above the W pair production threshold, new
phenomenological parameters  describing the TGV's become
relevant.   The electroweak chiral lagrangian provides a comprehensive
description of all these effects, coming from physics above ${\Lambda}_{\chi}$.

 All the electroweak chiral lagrangian operators up to dimension
 four have been included here, including one new CP invariant operator
 that has not been considered previously
 in the literature. This operator corresponds to the $g_5^Z$  term in the
 general parameterization of the TGV and gives a contribution linear in the
 mass difference of the heavy fermion doublet in the model used in our
 computation. This is in contrast to the quadratic dependence of
 $\beta_1$, $\alpha_8$ and $\alpha_9$
 on the mass splitting. On the other
 hand, as a consequence of $U(1)_{em}$ gauge invariance,  there is no
  corresponding chiral lagrangian operator for $g_5^{\gamma}$.

  The one-loop computation involving  one
flavor-doublet of heavy fermions (U and D) that carry technicolor number $N$
(but with the technicolor interactions neglected)
 indicates that the size of the TGV parameters $g_1^Z-1$,
 $\kappa_Z-1$ and
$\kappa_{\gamma}-1$ is of order $10^{-3}$ for $N=4$. In models
with more than one flavor-doublet, they could be of order
$10^{-2}$.  Assuming these results, with technicolor interactions neglected,
 are reliable estimates, they suggest that the
deviations from standard model predictions for the TGV parameters
 are within the reach of  a 500GeV $e^{+}e^{-}$ linear
collider, but beyond the precision of LEP-II.
 The size of $g_5^Z$  is of order
 $10^{-4}$ or  smaller in the one flavor-doublet model with $N=4$, depending
on the splitting of the masses of the heavy fermion doublet.

  All the CP violating operators of dimension-four
have also been listed in the appendix.  These include five new operators
that had not been listed. In most models, CP violating effects are
small, and we have not considered these effects in this paper.

\section*{Acknowledgments}
  We would like thank W. Marciano, S. Rey, J. Terning and G. Triantaphyllou
for helpful discussions. We are especially grateful to Martin Einhorn for
a critical reading of the manuscript and for several helpful suggestions.
This research is supported in part by a grant from the Texas National
Research Laboratory Commission.

\section*{Appendix A}
\setcounter{equation}{0}
\setcounter{section}{1}
\renewcommand{\theequation}{\Alph{section}.\arabic{equation}}

  All chiral lagrangian operators up to dimension-four satisfying
$SU(2)_L \times U(1)_Y$ gauge invariance, both CP conserving and CP violating,
are constructed in this appendix.
 By invoking the basic building blocks $T$, $V_{\mu}$ and $W_{\mu\nu}$
defined at the beginning of Section 2, the following list of operators with
 increasing dimension can be written down straightforwardly:

\vskip 0.1in
\begin{tabular}{|l|l|l|l|} \hline
 Dimension 0 &  $T$   &        &      \\   \hline
 Dimension 1 & ${\cal D}_{\mu}T$ &  $V_{\mu}$ &  \\  \hline
 Dimension 2 & ${\cal D}_{\mu}{\cal D}_{\nu}T$  &  ${\cal D}_{\mu} V_{\nu}$ &
 $W_{\mu\nu}$   \\    \hline
 Dimension 3 & ${\cal D}_{\mu}{\cal D}_{\nu}{\cal D}_{\rho}T$   &
 ${\cal D}_{\mu}{\cal D}_{\nu} V_{\rho}$ &  ${\cal D}_{\mu} W_{\nu\rho}$
\\ \hline
 Dimension 4 & ${\cal D}_{\mu}{\cal D}_{\nu}{\cal D}_{\rho}{\cal
D}_{\lambda}T$ &
 ${\cal D}_{\mu}{\cal D}_{\nu}{\cal D}_{\rho} V_{\lambda}$  &
 ${\cal D}_{\mu}{\cal D}_{\nu} W_{\rho\lambda}$    \\  \hline
\end{tabular}

\vskip 0.1in

The left covariant derivative ${\cal D}_{\mu}$ appearing in the above list is
 defined as:
\begin{eqnarray}
{\cal D}_{\mu} {\cal O} \equiv \partial_{\mu} {\cal O} + ig[W_{\mu}, {\cal O}]
\end{eqnarray}
where ${\cal O}$ is any  $SU(2)_L$ covariant and $U(1)_Y$ invariant operator.

 However, the above operators  are not all independent because of the following
identities:
\begin{eqnarray}
 {\cal D}_{\mu} T  =  [V_{\mu}, T]  \;\;\;\;\;\;\;\;\;\;\;
 [{\cal D}_{\mu},{\cal D}_{\nu}] {\cal O}  =  [W_{\mu\nu},{\cal O}]  \\
 {\cal D}_{\mu} V_{\nu} - {\cal D}_{\nu} V_{\mu}  =  igW_{\mu\nu} - i g'
 B_{\mu\nu}T + [V_{\mu},V_{\nu}]
\end{eqnarray}
For the purpose of applying the chiral effective lagrangian, the
equations of motion for the operator fields should also be properly taken
into account.  The equation of motion for the $V(x)$ field can be found to be:
\begin{eqnarray}
{\cal D}_{\mu} V^{\mu} \simeq 0
\end{eqnarray}
where the approximation is valid as long as the mass scale of the external
 fermions lies  well below the electroweak symmetry breaking scale
 $f\simeq 250GeV$, which is  true for the process $e^{+}e^{-} \rightarrow
W^{+}W^{-}$.

  With the above identities and the equation of motion, operators of the form
${\cal D}_{\mu}{\cal D}_{\nu} \cdots {\cal D}_{\rho}T$ and
${\cal D}_{\mu}{\cal D}_{\nu} \cdots {\cal D}_{\rho}V_{\lambda}$
 can be eliminated in favor of $T$, $V_{\mu}$ $W_{\mu\nu}$ and $B_{\mu\nu}$.
 While operators of the form ${\cal D}_{\mu}\cdots {\cal D}_{\nu}
 W_{\rho\lambda}$ can be gotten rid of via integration by parts.
We are left with only operators containing no ${\cal D}_{\mu}$ in them.
 After making use of the  trace identities of
 Pauli matrices ${\tau}_i$ and the fact that $T$, $V_{\mu}$ and $W_{\mu\nu}$
all can be expressed in terms of the linear combinations of ${\tau}_i$,
 only two independent trace structures exist:
\begin{eqnarray}
 & & Tr({\cal O}_1 {\cal O}_2)Tr({\cal O}_3 {\cal O}_4) \cdots
 Tr({\cal O}_{2n-1} {\cal O}_{2n})     \\
 & & Tr({\cal O}_1 {\cal O}_2)Tr({\cal O}_3 {\cal O}_4) \cdots
Tr({\cal O}_{2n-1}{\cal O}_{2n}) Tr({\cal O}_{2n+1} {\cal O}_{2n+2}
{\cal O}_{2n+3} )
\end{eqnarray}
where ${\cal O}=T$, $V_{\mu}$ or $W_{\mu\nu}$. With
$T$, $V_{\mu}$, $W_{\mu\nu}$ and $B_{\mu\nu}$, the
  chiral lagrangian operators of  dimensions two and four can be written down
systematically. The $CP$ invariant ones are given in section 2, the $CP$
noninvariant ones are:
\begin{eqnarray}
{\cal L}_{12} & \equiv & \alpha_{12} g Tr(TV_{\mu}) Tr(V_{\nu}W^{\mu\nu})  \\
{\cal L}_{13} & \equiv & \alpha_{13} g g' {\epsilon}^{\mu\nu\rho\sigma}
B_{\mu\nu}Tr(TW_{\rho\sigma}) \\
{\cal L}_{14} & \equiv & i \alpha_{14} g' {\epsilon}^{\mu\nu\rho\sigma}
B_{\mu\nu} Tr(T [V_{\rho}, V_{\sigma}]) \\
{\cal L}_{15} & \equiv & i \alpha_{15} g{\epsilon}^{\mu\nu\rho\sigma}
Tr(W_{\mu\nu} [V_{\rho}, V_{\sigma}]) \\
{\cal L}_{16} & \equiv & \alpha_{16} g^2 {\epsilon}^{\mu\nu\rho\sigma}
Tr(TW_{\mu\nu}) Tr(TW_{\rho\sigma}) \\
{\cal L}_{17} & \equiv & i \alpha_{17} g{\epsilon}^{\mu\nu\rho\sigma}
Tr(TW_{\mu\nu}) Tr(T[V_{\rho}, V_{\sigma}]) \\
{\cal L}_{18} & \equiv & \alpha_{18}
g'^2{\epsilon}^{\mu\nu\rho\sigma}B_{\mu\nu}
B_{\rho\sigma} \\
{\cal L}_{19} & \equiv & \alpha_{19} g^2 {\epsilon}^{\mu\nu\rho\sigma}
Tr(W_{\mu\nu}W_{\rho\sigma})
\end{eqnarray}
 Note that ${\cal L}_{12}$ above corresponds to ${\cal L}_{14}$ in
 Longhitano's list. And ${\cal L}_{13}$ through ${\cal L}_{17}$ have not been
listed before.


\newpage
\begin{thebibliography}{10}

\bibitem{AB}
T. Appelquist and C. Bernard, Phys. Rev. D22 (1980) 200.

\bibitem{Long}
A. Longhitano, Phys. Rev. D22 (1980) 1166; Nucl. Phys. B188 (1981) 118.

\bibitem{ASA}
T. Appelquist, in ``Gauge theories and experiments at high energies", ed. by
K.C. Brower and D.G. Sutherland,  Scottish University Summer School in
 Physics, St. Andrews (1980).

\bibitem{ewprecision}
For a recent review on this subject, see the proceedings of the
``Yale Workshop on Future Colliders", Oct. 2-3, 1992, to appear as a SLAC PUB.

\bibitem{Hol}
B. Holdom, Phys. Lett. B258 (1991) 156.

\bibitem{Ein}
M. Einhorn and J. Wudka, talk presented at the Workshop on Electroweak
Symmetry Breaking, Hiroshima, 1991, Santa Barbara Institute for Theoretical
Physics preprint, NSF-ITP-92-01, 1992.

M. Einhorn and J. Wudka, ``Anomalous Vector Boson Couplings--Fact and Fiction",
University of Michigan preprint, UM-TH-92-25, 1992.

\bibitem{BDV}
J. Bagger, S. Dawson and G. Valencia, Phys. Rev. Lett. 67 (1991) 2256 and
Fermilab-Pub-92/75-T, 1992.

\bibitem{FLS}
A. Falk, M. Luke and E. Simmons,  Nucl. Phys. B365 (1991) 523.

\bibitem{Ruj}
A. De R\'{u}jula, M.B. Gavela, P. Hern\'{a}ndez and E. Mass\'{o},
 Nucl. Phys. B384 (1992) 3.

\bibitem{BuL}
C. Burgess and D. London, Phys. Rev. Lett. 69 (1992) 3428.

\bibitem{EsH}
D. Espriu and M.J. Herrero, Nucl. Phys. B373 (1992) 117.

\bibitem{HeV}
P. Hernandez and F. Vegas, ``One Loop Effects of Non-Standard Triple Gauge
Boson
Vertices", CERN preprint, CERN-TH 6670, Dec. 1992.

\bibitem{Fer}
This term was also obtained by F. Feruglio, ``The Chiral Approach to
 Electroweak Interactions", Lectures given at the second National Seminar
of Theoretical Physics, Parma, Italy, 1-12 September 1992, DFPD 92/TH/50,
September 1992. Our explicit results for the TGV parameters
[see Section 4], however, differ from those of Feruglio. This is because we
have included the effects of the gauge boson two-point function renormalization
on the TGV parameters.

\bibitem{PeT}
M.E. Peskin and T. Takeuchi, Phys. Rev. Lett. 65 (1990) 964;  Phys. Rev. D46
(1992) 381.

\bibitem{Hag2}
K. Hagiwara, R.D. Peccei, D. Zeppenfeld and K. Hikasa, Nucl. Phys. B282
 (1987) 253.

\bibitem{Pes}
M. Peskin,  lectures presented at the 17th SLAC Summer Institute (1989),
SLAC report SLAC-PUB-5210, 1990.

\bibitem{Hol2}
For the subset of operators not containing weak isospin violation, these
connections were obtained by B. Holdom. See ref. \cite{Hol}.

\bibitem{WZW}
J. Wess and B. Zumino, Phys. Lett. 37B (1971) 95; E. Witten,
Nucl. Phys. B223 (1983) 422.

\bibitem{Cha}
M. Chanowitz, M. Furman and I. Hinchliffe, Nucl. Phys. B159 (1979) 225; B.~A.~
Ovrut, Nucl. Phys. B213 (1983) 241.

\end{thebibliography}
\end{document}